\documentstyle[aps,prb,twocolumn]{revtex}
\input{epsf.tex}
\input psfig.sty
\include{psfig}
\begin{document}
\draft
\title{Temperature dependence of the diffuse scattering fine structure
in equiatomic CuAu}
\author{O. Malis, K. F. Ludwig, Jr.}
\address{Boston University, Boston, MA.}
\author{W. Schweika}
\address{Forschungszentrum Juelich, Germany}
\author{G.E. Ice, C.J. Sparks}
\address{Oak Ridge National Laboratory, Oak Ridge, TN.}
\date{\today}
\maketitle
\begin{abstract}
The temperature dependence of the diffuse scattering fine 
structure from disordered equiatomic CuAu was 
studied using {\it in situ} x-ray scattering.
In contrast to Cu$_3$Au the diffuse peak splitting in CuAu
was found to be relatively insensitive to temperature.
Consequently, no evidence 
for a divergence of the antiphase length-scale at the
transition temperature was found. At all temperatures studied 
the peak splitting is smaller than the value
corresponding to the CuAuII modulated phase. 
An extended Ginzburg-Landau approach 
is used to explain the temperature dependence of the diffuse peak profiles in the ordering and
modulation directions.
The estimated mean-field instability point is considerably lower than is the case for Cu$_3$Au. 

\end{abstract}


The existence of fine structure associated with diffuse scattering 
peaks in disordered metallic alloys has been known from diffraction 
studies for over thirty years.\cite{sato62}
Recently, however, Reichert et al.\cite{Cu3Au} carried out the first {\it in situ} 
study of the temperature dependence of the diffuse scattering fine structure
in a disordered alloy -- Cu$_3$Au.  Their evidence for an unexpected
divergence of the antiphase domain separation in the disordered Cu$_3$Au alloy at
the first-order transition has prompted renewed interest in the phenomenon.
Reichert et al. concluded\cite{Cu3Au} that Fermi surface-induced effects\cite{Moss} could 
not solely account for such behavior and suggested that entropic effects must 
be important. This experiment triggered at least two  
theoretical attempts to explain the temperature dependence of the diffuse 
splitting. Ozolins et al. \cite{Ozolins} were able to predict a
temperature dependence of the splitting using a cluster expansion approach and 
attributed it to simple entropic effects.  Another theoretical approach due
to Tsatskis \cite{Tsatskis} attributes the behavior of the splitting to the 
temperature- and wavevector-dependence of the self-energy.  
However, the detailed accuracy of the theoretical models remains unclear.  
Tsatskis' high temperature expansion is not expected to be valid near the transition
point, and his $\alpha$-expansion itself requires the input of short-range order 
(SRO) parameters to calculate the diffuse scattering.
Wolverton et al.\cite{Wolverton} predict a peak symmetry that differs in detail
from that actually observed in Cu$_3$Au.
Moreover, for CuAu, they predict a splitting 
of the (100) diffuse peak along the [100] direction, which is in 
contradiction with published x-ray\cite{Hash83} and electron\cite{sato62} measurements.
Clearly further experimental and theoretical investigations are required to
resolve these outstanding issues.


This paper presents an {\it in situ} investigation of the temperature dependence of the 
diffuse scattering from equiatomic CuAu. 
Unlike Cu$_3$Au, CuAu exhibits a stable one-dimensional long-period superlattice
phase designated CuAuII between 385$^\circ$C and 410$^\circ$C.  
The CuAuII phase consists of an array of periodic antiphase boundaries 
with an average modulation wavelength ten times the size of the underlying 
unit cell.  Since CuAuII is not the ground state,
entropic effects are expected to play a major role in stabilizing the modulated 
phase and presumably in determining the nature of equilibrium fluctuations in 
the disordered phase.  


A splitting of the diffuse peaks in disordered 
equiatomic CuAu was reported in early electron diffraction studies\cite{sato62} and 
investigated in detail by Hashimoto\cite{Hash83} with x-rays. 
Hashimoto, however, performed his experiment on quenched samples at room 
temperature and therefore was not able 
to study the structure of the disordered phase in equilibrium or to examine its
temperature dependence.  Moreover, given the large atomic mobilities 
in the alloy at high temperature, it is possible, indeed likely, that 
his samples underwent significant ordering during the quench.  This possibility
is supported by a recent time-resolved x-ray study of ordering kinetics in CuAu
which is reported elsewhere.\cite{kin} 
The relatively rapid initial ordering kinetics that we observe in the alloy
suggests that accurate measurements of the diffuse scattering must
be performed at high temperature.

The diffuse x-ray scattering measurements reported here were performed at the 
National Synchrotron Light Source, Brookhaven National Laboratory. 
The samples studied were (421) and (100) cuts of a CuAu single crystal 
ingot grown by Monocrystals Inc. The composition of the (421) cut
was measured using fluorescence analysis and found to be 50.5$\pm$0.2 
at.\% Cu. During the experiment the samples were kept at temperatures
between 410$^\circ$ and 600$^\circ$C in 
vacuum or in a high-purity He atmosphere. 
Most of the x-ray data was taken on beamline X14 as part of an anomalous diffuse 
scattering project\cite{DIFF} and measured at three x-ray energies, 8959eV, 10500eV and 
11914eV.  Data at all three energies displays the same trends.  A 
mosaic, sagittal-focusing graphite (002) crystal spectrometer was used to separate 
the elastic signal from the resonant Raman by distributing it on a linear 
position-sensitive proportional counter.\cite{FeNi} The contribution 
due to Compton scattering is small and has not been removed from the data. 
The HWHM resolution of this experimental setup is better than 0.04 reciprocal 
lattice units, r.l.u. More information about this setup can be found in 
Ref. \onlinecite{FeNi} and references therein. Additional x-ray data on the
(100) diffuse peaks was taken on beamline X20C using a multilayer monochromator 
and CCD array detector. The resolution of this setup is 0.01 r.l.u.

All measurements of the diffuse scattering fine structure in Cu$_x$Au$_{1-x}$
alloys have shown a qualitatively similar symmetry of the scattering due 
to SRO (see Fig. 1 of Ref. \onlinecite{Cu3Au}). 
Around the (100) superlattice position 
the scattering due to SRO reaches its maximum in a ring 
in the (1$k$$l$) plane and has satellites on the [1$k$0] and [10$l$] 
lines.
These satellites are due to medium-range in-plane anti-phase
correlations between fluctuations ordering in the (100) direction and have a structure
reminiscent of the CuAuII modulated phase. The ($h$00) cross-section of the diffuse peak
reflects the degree of order in the [100] direction and will be referred to
as the ordering direction. The (1$k$0) and (10$l$) cross-sections probe the
degree of in-plane correlations and will therefore be referred to as modulation directions.
The scattering 
around other reciprocal lattice points can be derived using symmetry arguments. 

\begin{figure}[h]
\begin{center}
\mbox{\psfig{figure=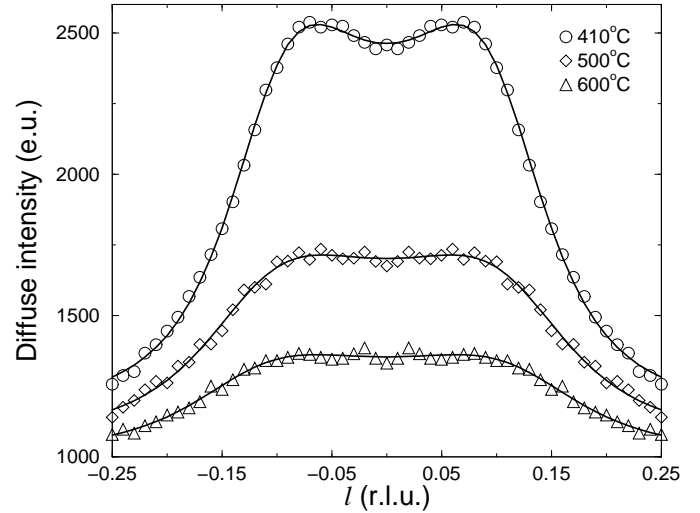,width=3.5in}} \end{center}
\caption{Diffuse scattering due to SRO along the (21$l$) line 
at three temperatures above the order-disorder
transition temperature of CuAu.}
\label{scatter}
\end{figure}

Our experiment agrees with earlier investigations on the symmetry of the diffuse
satellites. In addition to the SRO scattering, CuAu also 
exhibits strong size effects which shift the peaks considerably from their 
symmetric positions around the superlattice positions. Figure \ref{scatter} 
presents $l$ scans of the diffuse scattering due to SRO around the (210) peak at three 
different temperatures. With increasing temperature the satellite peaks
broaden to form a single peak with a relatively flat top.
The size-effect contribution to scattering on this line is minimal and has been
removed making use of its antisymmetry with respect to $l=0$.\cite{Borie} 
The data has been normalized by comparison with scattering from a powder nickel 
standard. The intensity of the diffuse peaks is at least an order of magnitude 
smaller that reported by Hashimoto, therefore confirming our concern that his 
sample underwent significant ordering during the quench.

In a simple mean-field approach the equilibrium fluctuations in the
disordered phase produce diffuse scattering peaks that have a Lorentzian
${\bf q}$-dependence. The HWHM of this Lorentzian is equal to the inverse of the 
correlation length, $\xi$, and depends on temperature as $(T-T_0)^{\nu}$. 
$T_0$ is the instability (spinodal) temperature and $\nu=1/2$ is the mean-field 
correlation length exponent.
The diffuse satellites we observe are characterized by two correlation lengths, 
one in the ordering direction and another in the modulation direction. 
At 410$^\circ$C the correlation length in the modulation direction, approximately 12$\overcirc{A}$,
is comparable in size to half the modulation wavelength and is almost 
four times larger than the correlation length in the ordering direction. 

\begin{figure}[h]
\begin{center}
  \mbox{\psfig{figure=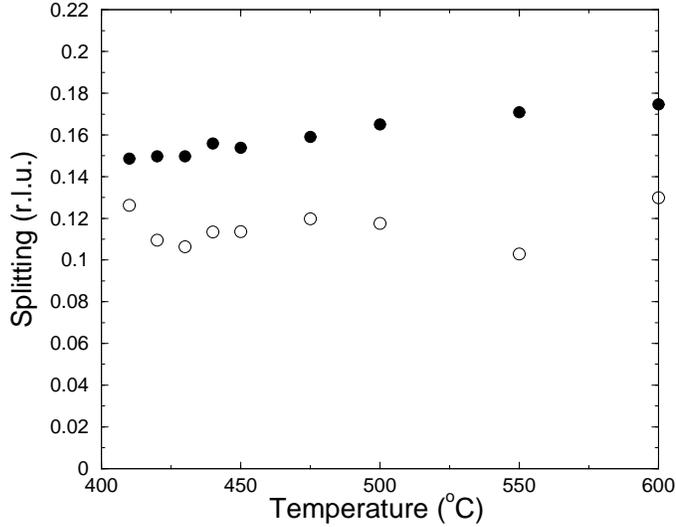,width=3.5in}} \end{center}
\caption{Temperature dependence of the CuAu fine structure diffuse scattering 
splitting from a fit of the scattering in the modulated direction with two 
Lorentzians (filled circles) and a fit with Eq.\ (\ref{fit}) (open circles).}
\label{split}
\end{figure}

In order to estimate the diffuse peak splitting in the modulation direction,
we fit the satellite peaks with individual Lorentzians plus a constant background. 
The fit also allows for the contribution of the two off-axis satellite peaks. 
Figure \ref{split} shows the temperature 
dependence of the peak separation $\delta l$.  The splitting decreases 
with decreasing temperature, but the magnitude of the overall change is small,
less than half the change observed in Cu$_3$Au\cite{Cu3Au} over a comparable 
temperature range, and may be an artifact of the fitting procedure.
There is no visible evidence of a diverging antiphase length-scale at the 
transition temperature. It is noteworthy that at all 
temperatures studied the values of $\delta l$ are smaller than 0.2 r.l.u., the 
value expected from the modulation wavelength of the ordered CuAuII phase. At 410$^\circ$C,
for example, the measured $\delta l$ corresponds to a modulation wavelength of 
approximately 13 unit cells as compared to 10 unit cells in CuAuII. 

\begin{figure}[h]
\begin{center}
  \mbox{\psfig{figure=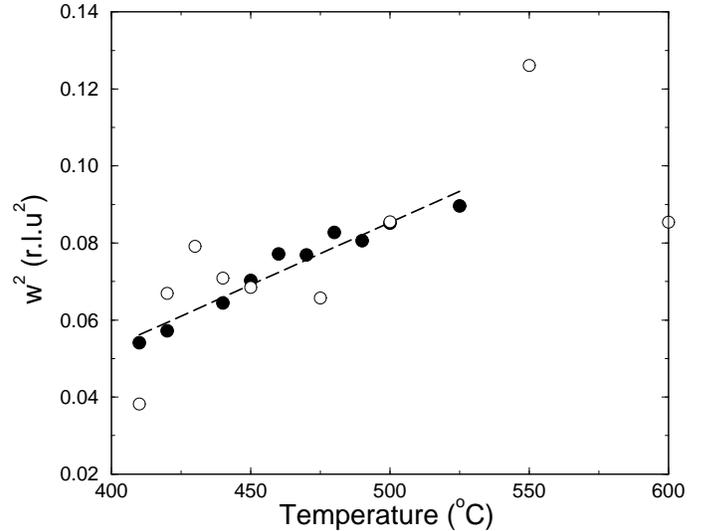,width=3.5in}} \end{center}
\caption{Temperature dependence of the square widths of the (100) diffuse peaks
measured in the ordering direction (filled circles) and calculated from the
$m$ and $n$ parameters obtained from the fit of the satellites in the modulated 
direction with Eq.\ (\ref{fit}) (open circles).
If a straight line is fit through the experimental points it extrapolates to a 
pseudo-spinodal temperature at approximately 240$^\circ$C.}
\label{spinodal1}
\end{figure}

We have studied in detail the temperature dependence of the two
correlation lengths characterizing the diffuse peaks.
In the modulation direction, the widths of the (210) satellite peaks obtained 
from the fit mentioned above show little change with temperature and 
would suggest a very low pseudo-spinodal point at approximately 150$^\circ$C.
In the ordering direction, however, the peak widths display a much 
stronger temperature dependence. Figure \ref{spinodal1} presents the temperature 
dependence of the square of the (100) diffuse peak widths in the ordering direction.
A straight line fitted through the experimental points would extrapolate to a 
pseudo-spinodal temperature located at approximately 240$^\circ$C. 

The contradiction between the behaviors of the correlation lengths in the ordering and modulation
directions would suggest that the
fitting of the satellite peaks with two individual Lorentzians is too simplistic 
to account for the effects associated with the anti-phase correlations. 
Another approach proposed to describe such correlations uses
a Ginzburg-Landau free energy functional with a negative 
gradient term:\cite{Landau2,Landau1}

\begin{equation}
F(\eta) = F_0 + {1 \over V}\int dr[a_T \eta^2 - e |\nabla_{mod}\eta|^2 + \\
\end{equation}
$$f(\nabla_{mod}^2 \eta)^2 + e |\nabla_{ord}\eta|^2]$$
where $\eta$ is the order parameter, $a_T=a(T-T_0)$, $F_0$, $a$, $e$ and $f$ are 
positive parameters approximately independent of temperature and $V$ is the volume.

The structure factor then can be written as

\begin{equation}
I({\bf q})=({{a}\over{2\pi}})^3{{k_b T/2}\over{a_T-e(q_{mod}^2-q_{ord}^2)+fq_{mod}^4}}
\end{equation} 
where  $q_{mod}$ is the component of the wavevector in the modulation direction
and  $q_{ord}$ is the component in the ordering direction.

In the ordering direction this model predicts that at $q_{mod}=0$ the diffuse peak
has a Lorentzian shape with a width that goes to zero at $T_0$ ( $\approx$240$^\circ$C in our case) which
is the instability point of the disordered phase with respect to the
ordered phase. At the satellite peak $q_{mod}=\sqrt{e/2f}$ the width of the
Lorentzian in the ordering direction goes to zero at $T_1=T_0+e^2/4fa$ which is the instability point of the
disordered phase with respect to the modulated phase. 
In the modulation direction the satellites have a non-Lorentzian line shape but 
their peak width also goes to zero at $T_1$.

At each temperature we fit the diffuse satellites in the modulation direction
with the following form:
 
\begin{equation}
I({\bf q})= {{C}\over{m-nq_{mod}^2+q_{mod}^4}}+const
\label{fit}
\end{equation}
where $C=k_b T (a/2\pi)^3/2f$, $m=a(T-T_0)/f$ and $n=e/f$. $C$, $m$ and $n$ are
allowed to change with temperature. The peak splitting calculated from $n$
(fig. \ref{split}) is consistent with our 
previous findings that the diffuse splitting does not change significantly with
temperature. This functional form produces slightly smaller values for the
peak splitting.

\begin{figure}
\begin{center}
  \mbox{\psfig{figure=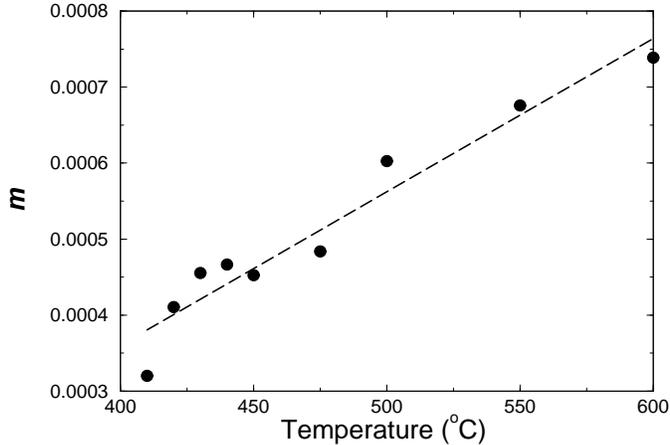,width=3.5in}} \end{center}
\caption{Temperature dependence of the $m$ parameter obtained from a fit
of the satellites in the modulated direction with Eq.\ (\ref{fit}). 
The straight line extrapolates to a pseudo-spinodal temperature approximately 
equal to 220$^\circ$C.
}
\label{spinodal2}
\end{figure}

Figure \ref{spinodal2} shows
the temperature dependence of $m$. The straight line extrapolates to 
$T_0\approx220^\circ$C which is in reasonable agreement with the
estimation of the pseudo-spinodal from the ordering direction.
Moreover, the values of the inverse square of the correlation length in the ordering
direction predicted from $m$ and $n$, though noisy, are in good quantitative agreement with the actual measured 
values as seen in figure \ref{spinodal2}. Thus equation \ref{fit} provides a good 
unified  description of the behaviors of the different correlation lengths in the
two perpendicular directions

\begin{figure}
\begin{center}
  \mbox{\psfig{figure=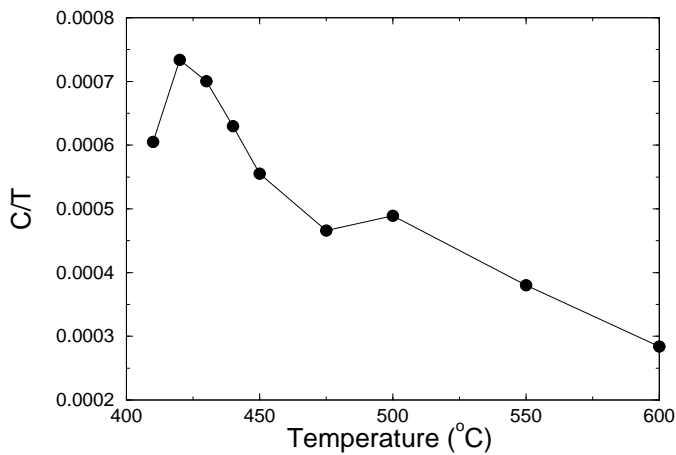,width=3.5in}} \end{center}
\caption{Temperature dependence of $C/T$ obtained from a fit
of the satellites in the modulated direction with eqn. 
}
\label{aT}
\end{figure}

Figure \ref{aT} shows the temperature dependence of $C/T$. We observe
a consistent deviation of $C/T$ from a constant. This increase in scattering 
with decreasing temperature cannot be explained by the simple Ginzburg-Landau 
approach and may be due to a non-critical increase of the number 
of atoms participating in these fluctuations beyond that expected from
simple mean-field spinodal theory.


In conclusion, we have presented a detailed {\it in situ} study of the diffuse
scattering fine structure in equiatomic CuAu. The satellite splitting is
smaller than in the equilibrium CuAuII phase and exhibits
little or no temperature dependence. 
In particular, there is no evidence of a divergence at the transition temperature.
The temperature dependence of the correlation lengths in the modulation and
ordering direction was explained coherently within the context of an extended Ginzburg-Landau
mean-field approach. The estimated pseudo-spinodal was found to be at least
170$^\circ$ below the ordering transition temperature, T$_{tr}$. This value, 
approximately 
0.25 T$_{tr}$, is considerably larger than the offset found in Cu$_3$Au
(0.05 T$_{tr}$).\cite{Cu3Au_spi}
This conclusion is not qualitatively changed if we 
instead use scaling exponents suggested by Monte Carlo simulations of 
CuAu.\cite{werner}

\section*{Acknowledgments}

The authors would like to thank K. Elder, B. Chakraborty and
W. Klein for numerous useful discussions. We would also like to acknowledge the 
experimental help of J. Bai. 
This work was supported by NSF grant DMR-9633596. The NSLS is supported by 
DOE Division of Materials Sciences and Division of Chemical Sciences.

\end{document}